\documentclass[aps,pra,twocolumn,superscriptaddress,showpacs,a4paper]{revtex4}

\usepackage{graphicx,amsmath,amssymb}

\begin{document}

\title{Optical amplification enhancement in photonic crystals}

\author{R. Sapienza}
\altaffiliation{Current address: ICFO-Institut de Ciencies
Fotoniques, Mediterranean Technology Park, 08860 Castelldefels (Barcelona), Spain}
\affiliation{Instituto de Ciencia de Materiales de Madrid (CSIC)
 and Unidad Asociada CSIC-UVigo, Cantoblanco 28049 Madrid
Espa\~{n}a.} \homepage[]{www.luxrerum.org}\email[]{cefe@icmm.csic.es}

\author{M. Leonetti}
\affiliation{Instituto de Ciencia de Materiales de Madrid (CSIC)
 and Unidad Asociada CSIC-UVigo, Cantoblanco 28049 Madrid Espa\~{n}a.}

\author{L. S. Froufe-P\'{e}rez}
\affiliation{Instituto de Ciencia de Materiales de Madrid (CSIC)
 and Unidad Asociada CSIC-UVigo, Cantoblanco 28049 Madrid Espa\~{n}a.}

\author{J. F. Galisteo-L\'{o}pez}
\affiliation{Instituto de Ciencia de Materiales de Madrid (CSIC)
 and Unidad Asociada CSIC-UVigo, Cantoblanco 28049 Madrid Espa\~{n}a.}

\author{C. Conti}
\affiliation{Research Center INFM-CNR, c/o Universit\'{a} di Roma Sapienza, I-00185, Roma Italy}

\author{C. L\'{o}pez}
\affiliation{Instituto de Ciencia de Materiales de Madrid (CSIC)
 and Unidad Asociada CSIC-UVigo, Cantoblanco 28049 Madrid Espa\~{n}a.}
\email[]{cefe@icmm.csic.es}
\date{\today}

\begin{abstract}
Improving and controlling the efficiency of a gain medium is one
of the most challenging problems of laser research. By measuring the
gain length in an opal based photonic crystal doped with laser
dye, we demonstrate that optical amplification
is more than twenty-fold enhanced along the $\Gamma$-K symmetry
directions of the face centered cubic photonic crystal. These
results are theoretically explained by directional variations of
the density of states, providing a quantitative
connection between density of the states and light amplification.

\end{abstract}


\maketitle
%---------------------------------------------------------------------------

The study of unconventional lasing is a novel and active field of
research. Mirror-less, micron-sized and three-dimensional lasers,
in the form of random lasers \cite{randomlasers}, micro-sphere and
microdrop lasers \cite{microcavity},  photonic crystal (PhC)
lasers \cite{PhClasers} and chaotic-cavity lasers \cite{caoticCav}
have been recently proposed as  more sophisticated alternatives to
standard lasers made of macroscopic mirrors and Fabry-P\'erot
cavities.

Artificially engineered photonic materials, with nanoscopic
features and topologies ranging from order to disorder, allow for
a superior control of light modes, dispersion and gain
\cite{ceferev}. Photonic structures can control and tune the
directionality and spectral extent of the emitted light without
relying on light absorption, allowing for efficient light sources
with smarter functionalities.

While the research of  2D PhC lasing can exploit high-quality and
low volume micro-cavities \cite{2Dlasing}, in three dimensions
(3D) PhC lasing is so far only based on the large light-matter
interaction that can be achieved by extended Bloch modes, at
band-edges \cite{PhClasingBedge} or at high-energy flat bands
\cite{PhClasingHigh}. These modes effectively  confine the light
in the active medium and increase the amplification and the gain
extraction from large bulk volumes \cite{PhClasingPrediction}. A
similar confinement, but originated by multiple scattering, is
responsible for random lasing of diffusive modes
\cite{randomlasers}.
The first pioneering study of gain in PhC has demonstrated a directional
modification of the gain spectral profile, attributed
to band-edge effects  \cite{gaininPhC}.  Enhanced optical gain was
also extrapolated from a decreased lasing threshold at the
$L$-pseudogap edges of an opal-TiO$_2$ composite containing quantum dots
\cite{gaininPhC2}.

    \begin{figure}[!h]
    \begin{center}
    \includegraphics[width=8cm]{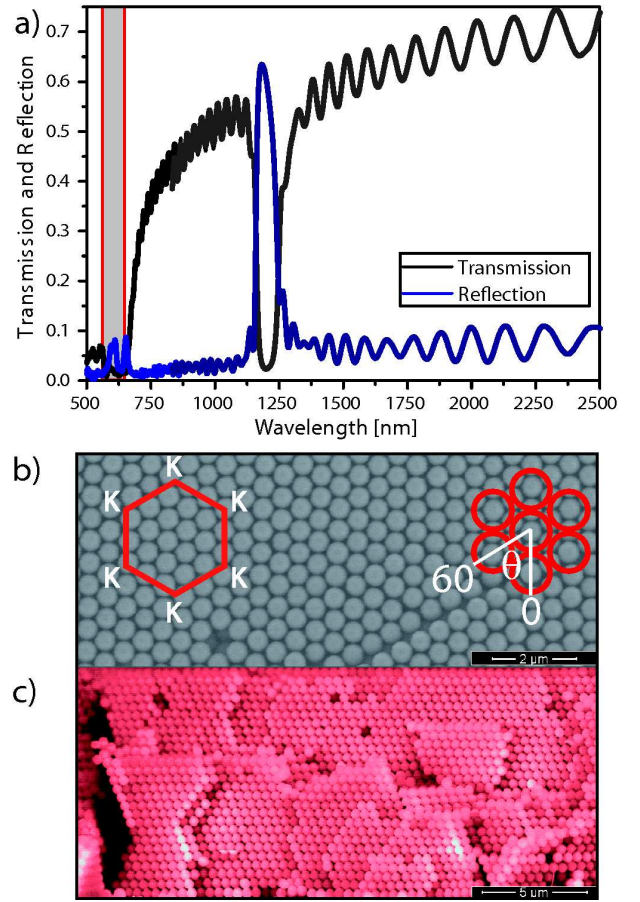} %figsetup}
    \caption{ a) Photonic crystal transmission and reflection spectra. The shaded region indicates the gain curve of the dye. b) SEM image of the top of the sample with a sketch of the
    crystallographic directions and the real space arrangement. The hexagon indicates a cut of 3d Brillouin  in the plane  (111)  c) SEM image of the side of the sample, corrected with the real color of the dye doped spheres. }\label{Sample}
    \end{center}
    \end{figure}

The lasing wavelength selection and the decrease of the lasing
threshold in PhC lasers has been often claimed to be connected to
spectral maxima of density of states \cite{PhClasingPrediction},
but a quantitative evidence of the directional nature of the gain
length due to the dependance of the density of the states on the
crystallographic direction in which the amplification grows has
never been reported so far.
Demonstrating this link is of paramount importance not only for
the future developments of nano-structured laser devices but also
as a starting point for several fundamental investigations ranging
from quantum information to slow light phenomena.

In this article we report a detailed measurement of the gain length
in opal-based photonic crystals doped by Rhodamine 6G. The
measured gain length has a six-fold symmetry, takes values up to
$\sim 500$ cm$^{-1}$ along the $\Gamma$-K directions in reciprocal
space and decreases to values below measurable levels of $\sim 20$
cm$^{-1}$ far from the high symmetry directions. We compare these
findings to the directional density of states which has been
predicted to be responsible for the increased gain.

The samples under study  are thin film artificial opals \cite{Colvin} consisting
of an fcc matrix of polystyrene spheres (PS, refractive index, $n = 1.59$, diameter $d = 512 \pm 15$ nm) bulk-doped with Rhodamine 6G with an emission  centered around  $\lambda = 574$ nm (from Duke Scientific). Figure \ref{Sample} shows transmission and reflection spectrum from the sample (in the panel a) as well as a two SEM pictures, of the top of sample (in panel b) and of its side (in panel c).
While a band-gap is visible at around 1200 nm, the dye emission indicated by the gray area sits in the high energy peaks of the spectrum around 574 nm.

\begin{figure}[!h]
    \begin{center}
 \includegraphics[width=8cm]{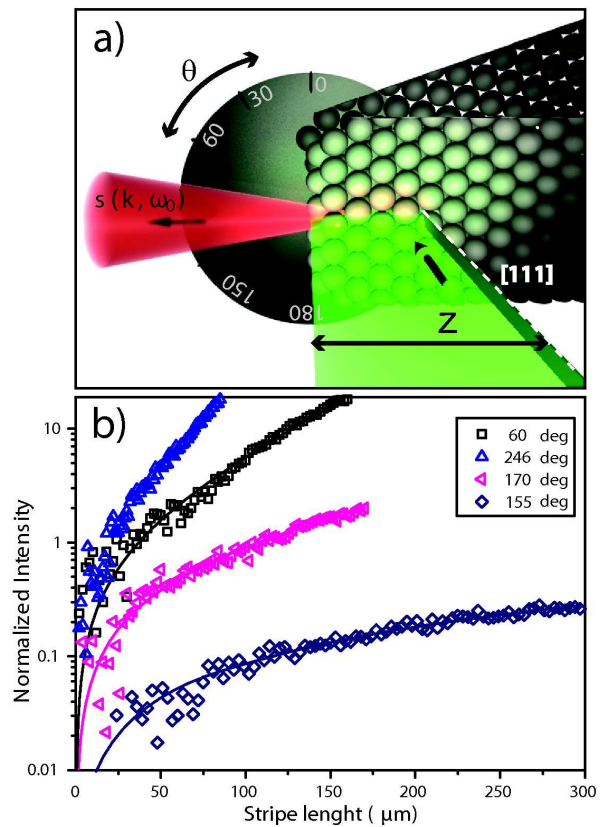} %figsetup}
    \caption{  (Color Online) a)  Sketch of two configurations for measurement of the $G$ parameter. The green laser
    pump is incident along the $\langle 111 \rangle$ direction and the red emission in the direction $\mathbf{s}\left(\mathbf{k},\omega_0 \right) $ is collected on a side by a miniature spectrometer. The PhC is rotated continuously in the plane $( 111 )$ orthogonal to the $[ 111 ]$ direction,  to explore the angle $\theta$ dependence of the emission. The illumination stripe has a variable length $z$. b) Intensity  of the emitted light, as a function of the length, collected at different angles. Black squares correspond to the emission
     at 60 deg, blue triangles at 246 deg, red triangles at 170 deg,  while blue squares indicate the angle of 155 deg. The continuous lines represent the fits obtained by using the equation reported in the experimental section. }\label{setup}
    \end{center}
    \end{figure}

The optical gain in the dye-doped three-dimensional artificial
opals can be extracted by employing the stripe length technique
\cite{stripelength,DNAgain}. The emission intensity and amplification is measured as
a function of the length of the illuminated area on the sample. In figure
\ref{setup}a, a sketch of the measurement
configurations used for the experiment is presented, while figure \ref{setup}b report the curves from which the amount of optical amplification is extracted.
The spontaneous emitted photons, traveling in the direction $z$ of the
illuminated stripe, undergo an exponential amplification of the
intensity with respect to the excitation length $z$ as they pass
trough an area of population inversion. This can be quantified by
the gain coefficient, the most important actor in light
amplification and lasing, which is a robust quantity, independent
on the sample thickness (in absence of pump depletion) and on the
edge coupling of the light from the stripe to the detector.
The output intensity $I(z)$ collected at one edge of the stripe
is the sum of spontaneous and stimulated emission and
has the well-known spatial dependence \cite{stripelength}:
\begin{subequations}\label{stripe}
\begin{eqnarray}
%I(z) =  I_{SE} \, [\exp(\frac{z}{ g-\ell_a^{-1}-\ell_s^{-1}})-1],\\
I(z) &=&I_{SE} \, [\exp{(G \, z)}-1]\\
G &=&g - \ell_a^{-1}-\ell_s^{-1},
\end{eqnarray}
\end{subequations}
where  $I_{SE}$ is the spontaneous emission (SE) collected by the
detector, $g$ is the gain (including confinement effects),
$\ell_a$ the absorption length and $\ell_s$ the scattering mean
free path. The quantity $G$, refereed to as the {\it net gain}, is
directly obtained from stripe-length measurements. Equation \ref{stripe}
a, holds until gain saturation is reached, which occurs for stripe
lengths larger than around 150-200 $\mu$m, in our case.

In our experiment the excitation is carried out by illuminating
the sample along the PhC $[ 111 ]$  direction (like in figure \ref{setup}a) with a Q-switched
Nd:Yag laser, 9 ns pulses, 10 Hz repetition rate, at a wavelength
of 532 nm.  The laser beam has the shape of a stripe of thickness
$\sim 20 \mu$m obtained by focusing the laser source with a
cylindrical lens of 75 mm focal. Measurements of the dependence of
$I$ as a function of $z$ has been performed by cutting the stripe
with a  blade moved automatically with micrometer accuracy. A
miniature fiber-coupled spectrometer is used to efficiently
collect the light along the stripe direction, as described
in ref. \cite{DNAgain}.

By rotating the stripe relative to the in-plane direction
$\mathbf{s}$, various crystallographic directions , including the
high-symmetry ones, can be explored. In Fig. \ref{setup}b, the
dependence of the collected fluorescence intensity on the
stripe length $z$, for various different orientations of the
photonic crystal is shown.
Emission for light propagating along two $\Gamma$-K crystal directions
is presented (60 and 240 deg), as well as an intermediate position (170 deg).
A non symmetrical direction is probed too (155 deg). These measurements show
a clear dependence of the fluorescence intensity on the propagating angle in the crystal.

\begin{figure}[!h]
    \begin{center}
 \includegraphics[width=8cm]{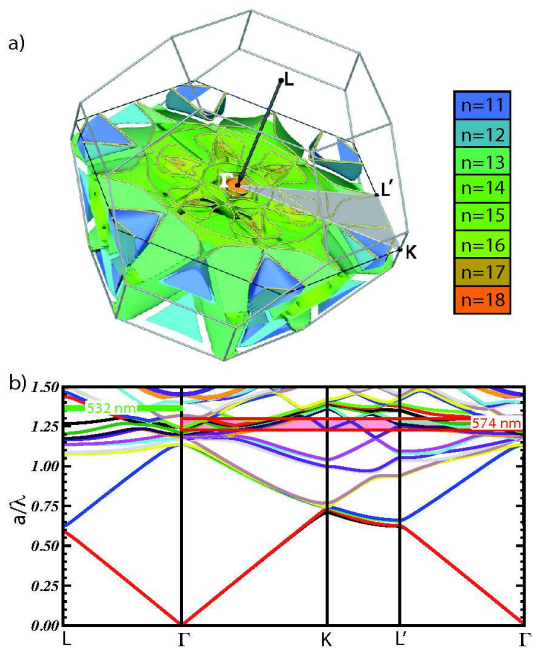}
    \caption{ \label{fig2} (Color Online) a) Isofrequency surface section in the plane,
    for  $\lambda =  574$ nm, cut in the plane $\perp$ to the  $\langle 111 \rangle$ direction,
    $\parallel$ to the sample surface, which is the plane of study. The color code refers to the band number indicated by $n$. b) Band diagram. The shaded red region indicates the gain curve of the dye centered at  $\lambda = 574$ nm, while the green line is the pump laser at $\lambda = 532$ nm.
    Therelevant symmetry directions are shown by the capital letters. }
    \end{center}
    \end{figure}

This result is a consequence of the photonic crystal complex band
structure that alter the optical gain and amplification processes
as compared to an homogenous medium.
In figure \ref{fig2}a we report the isofrequency surface
at the working frequency. For the sake of clarity only half the
isofrequency, cut along a plane perpendicular to the $[ 111 ]$
direction, is presented. Figure \ref{fig2}b shows the calculated band
structure along a path in the reciprocal space (sketched in panel
a). From the full 3D band structure (calculated by
preconditioned conjugate-gradient minimization of the block Rayleigh
quotient in a planewave basis, using a freely available software package
\cite{mpb}) the isofrequency surface and group velocities are obtained.

In the band diagram in figure  \ref{fig2}b, the directions $\Gamma-$K, K-L
and L'-$\Gamma$
 are explored when rotating the sample around the
direction  $[ 111 ]$ normal to its surface.
The shaded red region indicates the gain curve of the dye which
excites the high order diffractive modes when pumped by a
frequency doubled Nd:Yag laser (wavelength 532 nm) which is
indicated by the green line.
Note that $L$' is not a high symmetry point equivalent to $L$, but
just the midpoint of the experimental trajectory along the
hexagonal facet of the Brillouin zone.

An  homogenous media would exhibit a circular (resulting from the
section of a sphere) isofrequency contour; in a photonic crystal a
rich structure with a non-constant curvature is visible, which
results from contributions of eight different bands around the
emission energy. Each point in the isofrequency surface identify a
wave-vector $\mathbf{k}$, for which the direction of propagation
of the energy $\mathbf{s}$ is given by the gradient of the
surface.

   \begin{figure}[!h]
    \begin{center}
 \includegraphics[width=8.5cm]{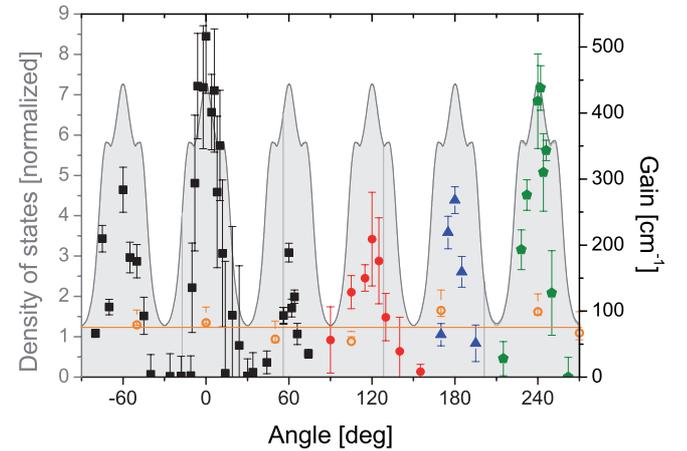}
    \caption{ \label{fig3}  (Color Online) Measured gain in a PhC as a function of the excitation angle. The orange thin line corresponds to     the homogenous reference sample.    The emission from the PhC is presented by the full symbols, while the gain of the reference sample is denoted by the empty orange circles. Different symbol styles refer to measurements performed after a 90 deg sample rotation.
    The curve above the gray shaded region denotes the DOS.  $\Gamma$-K directions occur each 60 deg from $\theta=0$ deg.}
    \end{center}
    \end{figure}

By fitting the data in figure \ref{setup}b to equation \ref{stripe} , the net gain $G$ can be extracted and plotted as a function of the stripe direction as in figure \ref{fig3}.  $G$ is affected by the variations of group velocities and varies as a function of the probed
crystallographic direction.
The angle $\theta = 0$ corresponds to the configuration in which the
stripe is aligned with the  opal growth axis, and it identifies a
$\Gamma$-K direction. A large variation of $G$ is visible with a
clear C6 symmetry and maxima along the six $\Gamma$-K directions
(each 60 deg from $\theta=0$ deg) reaching values up to 500
cm$^{-1}$. The minima measured along less-symmetrical directions
reach values which are limited to $\sim 20$ cm$^{-1}$ by the
measurement sensitivity:  a more than 20-fold variation is
observed. Different color symbols in figure \ref{fig3} corresponds to
measurement sets that have been performed  by rotating the sample
90 degrees. Each set of data is thus obtained on a different side
of the sample. The good quality of the latter - and, especially,
the homogeneous dye distribution - allows a high reproducibility
of the measurements, as shown in figure \ref{fig3}.
%
%Coupling effects due to the sample rotation do not influence the
%gain measured, as $G$ quantifies the increment of amplification in
%the $z$ direction and not its absolute value.

In order to prove that the observed effect depends on the regular
disposition of the spheres, a reference sample has been measured
too: the same PhC sample after melting by keeping at 100 Celsius  overnight.
This temperature is enough to
deform the PS spheres while keeping the film shape (as confirmed
by SEM inspection) without degrading the dye molecules. The
reference sample has a dye molecule density  35\% higher than the
regular sample, as it is a compact film as compared to the opal
PhC whose filling fraction is 0.74. The measured gain of 99 $\pm$
20 cm$^{-1}$ in the reference is now independent on the excitation
angle. The orange thin line in figure \ref{fig3} shows the reference
sample gain multiplied by 0.74 to account for the dye molecule
density increase.

The data are compared to the directional density of states $\rho\left ( \mathbf{s},
\omega \right )$ that  is calculated as the density of modes with
Poynting vector pointing in the stripe direction, given by
contribution of modes with any $k$-vector in the 3D space, and shown
in figure \ref{fig3} as the curve above the gray shaded region.

The extraordinary increase in optical amplification and  $G$ in figure
\ref{fig3}, as predicted by Sakoda
\cite{PhClasingPrediction,Sakodabook}, is connected to the
variation of the density of states $\rho\left ( \mathbf{s}, \omega
\right )$ in the direction $\mathbf{s}$ of the stripe. The number
of available optical states in a given direction, i.e. for a given
Poynting vector parallel to $\mathbf{s}$, determines the emission
probability for the excited dye molecules. A higher value of
$\rho\left ( \mathbf{s}, \omega \right )$ corresponds to a higher
emission probability in the $(\mathbf{s},\omega)$ mode and
therefore a larger gain for photons travelling in the $\mathbf{s}$
direction.
%
%The density of states $\rho\left ( \mathbf{s}, \omega \right )$ is
%also responsible for large directional variations of the light
%scattering strength in photonic crystals \cite{rapidPRB}.
The shaded region in figure \ref{fig3} shows the variation of $\rho\left (
\mathbf{s}, \omega \right )$ along the various directions in the
$( 111 )$  plane.
The theoretical curve in figure \ref{fig3} has been rescaled by the ratio
of the computed value of $\rho\left ( \mathbf{s}, \omega \right )$
for an homogenous medium to the measured gain in the reference
melted opal.

In general cases, also scattering and absorption can vary with the
angle. Far from the dye absorption curve, the opal absorption
length has been measured to be $\ell_a \sim 10$ m
\cite{sapienzaPRL}, which is therefore negligible for the system
under study. The scattering mean free path $\ell_s$ can present
indeed large variations with $\rho\left ( \mathbf{s}, \omega
\right )$, as studied in ref. \cite{rapidPRB}. For lower energy
and similar opals it has been measured to be in the range
$100-500$ $\mu$m. This value is larger but comparable to the gain
length $\ell_g = G^{-1}$ which is here in the range 20-200 $\mu$m.
Large scattering is anyhow responsible only for a decrease of the
amplification and cannot explain the large increase in $G$ that we
have measured. If the variations of $\ell_g$ were governed by
$\ell_s$, then the value of $\ell_s$ should be smaller than
$\ell_g$ and no net gain could be measured.
In our case scattering losses reduce the measured gain in figure
\ref{fig3} and set a minimum value of $G$ that can be measured in
our experiment ($\sim$20 cm$^{-1}$).

In conclusion we have presented an observation of a six-fold
increase of the gain in opal photonic crystal, as compared to the
homogeneous film, and a more than 20-fold variation of $G$ between
$\Gamma$-K and less symmetrical directions, in the same photonic
crystal. We explain this enhancement as due to a large increase of
the density of the available modes when exciting around the
$\Gamma$-K direction.
Large variations of the gain in PhC show the impact of the
tailored density of states on light generation and amplification
and open the way to enhance other phenomena like nonlinear wave
mixing and harmonic generation.
Our results show how nanostructured media could be at the basis of
the development of novel lasing sources with exceptional
tunability, directionality and efficiency while being plastic
photonics CMOS compatible, and candidates for in-board
interconnections for future generation computers.

%--------------------------------------------------------------------

\textit{Acknowledgement} We thank M. Ibisate, C. Soukoulis, K. Sakoda,
and J.J. Saenz for fruitful discussion. C.C. acknowledges the ERC GRANT
(FP7/2007-2013) n.201766. RS acknowledges support by RyC, LSFP
acknowledges  support by JdC, JFGL acknowledges support by JAE.
The work was supported by MAT2006-09062, the Spanish
MEC Consolider-NanoLight.es CSD20070046 and the Comunidad de Madrid
S-0505/ESP-0200.

%
%--------------------------------------------------------------------
%

%

\end{document}